\documentstyle[prb,aps,epsfig]{revtex}
\begin{document}

\draft
\flushbottom
\twocolumn[\hsize\textwidth\columnwidth\hsize
\csname@twocolumnfalse\endcsname

\title {
The Travelling Cluster Approximation for 
Strong Correlation Models of Lattice Fermions Coupled to Classical 
Fields}

\author{Sanjeev Kumar and Pinaki Majumdar }

\address{ Harish-Chandra  Research Institute,\\
 Chhatnag Road, Jhusi, Allahabad 211 019, India }

\date{June 3,  2004}

\maketitle
\tightenlines
\widetext
\advance\leftskip by 57pt
\advance\rightskip by 57pt
\begin{abstract}

We suggest and implement a new Monte Carlo strategy for correlated models
involving fermions strongly coupled to classical degrees of freedom, with 
accurate handling of quenched disorder as well. Current methods  
iteratively diagonalise the 
full Hamiltonian for a system of $N$ sites with computation time
$\tau_N \sim N^4$. This limits achievable sizes to $N \sim 100$. In 
our method 
the energy cost of a Monte Carlo update is computed from
the Hamiltonian of a cluster, of size $N_c$, constructed around the reference
site, and embedded in the larger system. As MC steps sweep over the system,
the cluster Hamiltonian also moves, being reconstructed at each site where
an update is attempted. In this method $\tau_{N,N_c} \sim NN_c^3$.
Our results are obviously exact when $N_c=N$,
and converge quickly to this asymptote with increasing $N_c$.
The accuracy  improves in systems where
the effective disorder seen by the fermions is large. We provide results
of preliminary calculations on the Holstein  model and the Double Exchange
model. The `locality' of the energy cost, as evidenced by our results, 
suggests that several important but inaccessible problems can now be handled
with control.

\

\

\end{abstract}

]

\narrowtext
\tightenlines

\section{Introduction}

The equilibrium physics of classical interacting systems is by now very
well understood. Quantum many body problems, involving strong interactions,
however, remain difficult to solve with control.
 The focus of strong correlation theory
is on  devising methods to handle these problems. Techniques like 
density matrix renormalisation group \cite{dmrg-ref} (DMRG) and dynamical
mean field theory \cite{dmft-ref} 
(DMFT), for example, represent an advance in this
direction. Problems where fermions are coupled to `classical fields',
{\it e.g}, large $S$ spins, or `adiabatic' phonons, 
are at an intermediate level of difficulty between
purely classical systems and quantum many body problems. 
The quantum degrees of freedom are not directly interacting, 
so the difficulty with an exponentially growing Hilbert space is 
absent, but ``annealing'' the classical variables is much more
difficult compared to purely classical systems.

The adiabatic approximation, whereby some degrees of freedom are
treated as classical,
is not novel.  Several problems have been solved in the past 
by making this approximation, {\it e.g}, 
in electron-phonon systems \cite{adiab-elphon}, or,
in a different context, in  the Car-Parrinello 
method \cite{car-parin}, handling coupled electronic and ionic 
degrees of freedom. The recent interest lies in the
application of this approach to several strong coupling
lattice fermion models, and some degree of success in
understanding complex materials.
Millis and coworkers \cite{mill-mull} studied  electrons 
coupled to classical spin and 
lattice (`phonon') degrees of freedom, to 
construct an initial theory of the manganites using
DMFT.
The approach was taken much farther by 
Dagotto and coworkers \cite{dag-ref}  using `real space' Monte Carlo techniques
to study ordering phenomena, phase coexistence, and disorder effects
in a large family of correlation models pertinent to the manganites.
The method has been used extensively also  
to explore  magnetism in double exchange (DE) based models \cite{de-mc}.  
For  diluted magnetic semiconductors (DMS) too  much of
the physics has been clarified by 
methods which treat the
doped magnetic moment  as classical \cite{dms-refs}.  
We ourselves have used the approach  
to study magnetism, insulator-metal transitions and 
nanoscale phase coexistence in 
disordered correlated electron models \cite{sk-pm-dde,sk-pm-nano}.

The adiabatic limit simplifies the many body problem by
casting it in the form of  ``non interacting'' fermions in the
background of classical variables, $\{x\}$ say,
but determining the distribution  $P\{x\}$ involves an 
expensive  computation. In the
absence of any small parameter to simplify the problem, 
computing $P\{x\}$ requires  iterative diagonalisation
\cite{ed-mc-ref} of
the fermion Hamiltonian and, for an $N$ site system, the
computation time, $\tau_N $, increases as
$ N^4$. 
We will describe the standard exact diagonalisation
based Monte Carlo (ED-MC) in the next section, here we just note
that the accessible sizes, $N \sim 100$, within ED-MC, severely
limits the ability of the method to resolve the outstanding issues
relating to transport, metal-insulator transitions, and the effect
of disorder in correlated systems.

There have  been some attempts at overcoming the severe finite size
constraint in ED-MC. $(i)$~Instead of exact diagonalisation, it has been
proposed that the energy of  fermions in the
classical background can be estimated by moment expansion of
the density of states. This,
in principle,  allows
access to $N \gtrsim 10^3$, and has been used to study the
clean \cite{furu-clean} and disordered \cite{furu-dis}
DE model. $(ii)$~A `hybrid'
Monte Carlo method, using dynamical evolution of the classical
variables, has been tried out \cite{hyb-mc}
for a model of competing DE 
and superexchange. $(iii)$~We have proposed a scheme \cite{sk-pm-scr},
in the context of double exchange,
where the energy associated with the spin configuration can
be approximated by an {\it explicit} classical Hamiltonian 
with couplings determined from a solution of the fermion 
problem.

While  
the approximations above have allowed some advance in the context of
double exchange, there is no equivalent method available
for handling phonon degrees of freedom, or the combination
of phonons and spins (as relevant to manganites), or dilute
strong coupling systems like the DMS.
There is the need for a {\it 
general and computationally transparent method}
that can handle models with arbitrary coupling and disorder,
and systematically approach the `exact' answer.
This paper proposes such a scheme. We employ  
a variant of the exact diagonalisation strategy 
using an embedded (travelling) cluster, that 
estimates the energy cost of a Monte Carlo move 
by diagonalising the smaller cluster rather than the full
Hamiltonian.
Since size limitations are most severe in three dimension ($3d$), and it is
physically also the most relevant, we benchmark our method directly
in $3d$, where the test is most stringent.

We will study two models, in $3d$, to provide some
performance benchmarks on the travelling cluster approximation (TCA). 
These are $(i)$~the 
(disordered) Holstein model, and $(ii)$~the double exchange
model. They are, respectively:
\begin{eqnarray}
H_1& =& -t\sum_{\langle ij \rangle}  
c^{\dagger}_i c^{~}_j 
+  \sum_{i } (\epsilon_i - \mu)   n_i  
- \lambda \sum_i n_i  x_i  + H_K \cr
H_2 &=&  -t\sum_{\langle ij \rangle, \sigma}  
c^{\dagger}_{i\sigma} c^{~}_{j \sigma}
+  \sum_{i } (\epsilon_i -\mu) n_i  -J_H \sum_i {\bf S}_i.{\hat \sigma}_i
\end{eqnarray}
The $t$ are
nearest neighbour hopping on a three  dimensional lattice.  
In $H_1$, 
the on site 
binary disorder $\epsilon_i$ assumes value $\pm \Delta$, 
 $\mu$ is the chemical potential and   
$n_i = c^{\dagger}_i c_i$ is the electron density operator
(for spinless fermions), 
coupling  to
the phonon coordinate $x_i$.
$H_K= (K/2)\sum_i x_i^2$ where
$K=1$ is the
stiffness of the phononic oscillators. In $H_2$,  $J_H$ is
the Hunds coupling. The 
model is defined with `spinfull' fermions, but since we will
use $J_H/t \rightarrow \infty$ 
it will also lead to an effective spinless fermion problem.
We set $t=1$, fixing our basic energy scale,
and also $\hbar =1$.

\section{ Method}

Let us start with the  $T=0$ case 
to clarify the usual approach to these problems.
Since we have earlier discussed the ED-MC method in detail \cite{sk-pm-scr}
we only provide a brief outline here.
There are several applications of ED-MC in the context of manganite related 
models \cite{ed-mc-ref}.

There are two (related) difficulties in solving strong coupling adiabatic
problems. $(i)$~The probability of occurence of a classical 
configuration is not explicitly known, and is governed by the fermion 
free energy. Generating these configurations involves the $N^4$ cost
specified earlier. Let us call this the ``annealing problem''.
$(ii)$~Even for a specified classical configuration, obtained via
some annealing technique, the electronic properties, {\it e.g}, the
resistivity, involves computing fermionic correlation functions in a 
non trivial background. Since there is no analytic theory 
for non interacting fermions in an arbitrary `landscape', transport
calculations have to implement the Kubo formulae exactly. Our innovation 
in this paper is on the annealing problem, we still depend on a
numerical implementation of linear response theory \cite{sk-pm-scr}
to solve the 
transport problem on large lattices.

Let us start with the annealing problem.
If we are at $T=0$, the background in which the fermions 
move can be determined by minimising the 
total energy $\langle H \rangle$ with respect to the classical
variables.
Denote the  classical configuration as 
$\{ \eta_1, \eta_2, ..\}
\equiv  \{\eta\}$,
where $\eta_i = \eta({\bf R}_i)$ and, in case of multiple classical
variables at each site, $\eta_i$ represents the full set of 
variables $x_i , {\bf S}_i$, {\it etc}, at that site.
The key task is to determine $\langle H \rangle ={\cal E}\{ \eta\}$.

If the coupling between the classical and quantum variables
is {\it large}, there is no perturbative result for 
the fermion energy 
${\cal E}\{ \eta\}$, 
and therefore 
no explicit functional that we can minimise.
This is where MC is used. The exact diagonalisation based
method uses the following strategy:
$(i)$~Set up an arbitrary configuration $\{ \eta \}$ and 
compute the energy 
${\cal E}\{ \eta\}$.   The fermion contribution is estimated 
by direct diagonalisation while the classical 
contribution, $Kx_i^2$ say, is explicit.
$(ii)$~Attempt an update, say at site ${\bf R}_i$, by
changing  $\eta_i \rightarrow \eta'_i$. Compute the energy 
${\cal E}\{ \eta'\}$.
$(iii)$~If  
$\Delta {\cal E} = {\cal E}\{ \eta'\} -
{\cal E}\{ \eta\} < 0$, accept the move, if 
$\Delta {\cal E} > 0$, and $T \neq 0$, accept the move with
probability $\propto e^{- \Delta {\cal E}/T}$.
$(iv)$~Sweep over the system, initially to reach equilibrium and
then to compute thermal averages.

The method above is simply a use of the Metropolis algorithm,
with the complication of an expensive diagonalisation for every
microscopic update. Since each local update involves computational
effort $\sim N^3$, the cost of sweeping over the system leads to
$\tau_N \sim N^4$. One has to multiply this with the cost of
thermal averaging, and disorder average (if needed).

We were motivated to ask if it is really necessary to diagonalise
the full Hamiltonian of the $N=L^3$ system to estimate the 
cost, $\Delta {\cal E}$, of a local move. 
Imagine a `large' system, 
(with $L=20$, say, for arguments sake) and
some degree of `disorder' seen by the electrons arising from the
classical thermal fluctuations or quenched disorder. Qualitatively, if
the effect of a change, $\eta_i \rightarrow \eta'_i$, does not
`propagate' very far, as one would expect in a system with some
disorder, then the energy cost of the move 
should be calculable
from a Hamiltonian which involves only 
electronic degrees of freedom
in the `vicinity' of ${\bf R}_i$. We will discuss the analytic
basis of such an argument separately, for the Monte Carlo 
it only requires 
that we modify step $(ii)$  of the ED-MC strategy, discussed earlier. 
We compute  
$\Delta {\cal E}$ as ${\cal E}_c\{ \eta'\} -
{\cal E}_c\{ \eta\}$, where 
${\cal E}_c\{ \eta\}$ is the energy computed by constructing a 
Hamiltonian of $N_c = L_c^3$ sites around ${\bf R}_i$, and diagonalising
this Hamiltonian in the background configurations $\{ \eta \}_c$ and
$\{\eta'\}_c$, where the curly brackets, $\{\}_c$,
refer to the configuration
within the cluster. 
We will show results on `equilibriation' within the TCA approach in the last
section.

After equilibriation the fermion properties are computed by diagonalising the
{\it full} $L^3$ Hamiltonian in the equilibrium background. Transport
properties are calculated based on the spectrum and states
obtained from these diagonalisation, using the Kubo formula \cite{sk-pm-scr}.

\section{ Results}

Apart from the electronic parameters, the two computational
parameters in the problem are $N$ and $N_c$. In our notation
TCA$(N : N_c)$ implies MC for a $N$ site system based on a cluster
of size $N_c$. ED-MC obviously corresponds to TCA$(N:N)$. We 
assume a cube geometry, with periodic boundary conditions (PBC)
 for both the system and the cluster.
Ideally one should have TCA$(N:N)$ available for large sizes
and study convergence as $N_c \rightarrow N$. Unfortunately
ED-MC can be done, with great effort, only for sizes 
$\lesssim 6^3$, so TCA$(N:N)$ will be rarely available at large $N$
and we have to analyse the approximation based on the following checks.

$(1).$~We study
$H_1$ and $H_2$ using ED-MC on the largest possible lattice,
$N = 6^3$.
With somewhat reduced thermal averaging, and using scan in
$T$ (at fixed electron density), or a scan in $\mu$
(at fixed temperature),
we establish the ``large size''
exact results.
We then use TCA$(N : N_c)$ with $N_c = 3^3, 4^3$ and $5^3$ 
to check the convergence to TCA$(N:N)$.

$(2).$~We study a ``large'' system, $N=8^3$,  and monitor the trend
in TCA$(N:N_c)$ with growing $N_c$, remaining in the regime
$N_c \ll N$. We  use 
$N_c = 3^3, 4^3, 5^3$ to assess 
the convergence of the results 
to an asymptote with  $N_c$ still $ \ll N$.

$(3).$~We compare the energy cost of actual microscopic MC updates
between ED-MC and TCA. We evolve a system via TCA$(8^3:4^3)$
but simultaneously compute the energy cost of the  updates via 
exact diagonalisation of the full $8^3$ system. This is done by
choosing a site randomly and computing both the exact $8^3$ energy 
cost and the $L_c^3$ energy cost whenever an update is attempted at
that site.
This yields information on how well TCA {\it microscopically}
estimates the energy cost, rather than at the level of system 
averaged properties.

\subsection{Clean Holstein Model}

The Holstein model provides the minimal description of coupled
electron and phonon degrees of freedom and, in the adiabatic
limit,  involves the following
phases \cite{dmft-holst1,dmft-holst2,dmft-holst3}, 
$(i)$~a Fermi liquid (FL) metal, without any lattice
distortions at $T=0$, $(ii)$~a positionally 
disordered insulating polaron liquid (PL)  
 at strong coupling, and $(iii)$~charge 

\begin{center}
\begin{figure}
\epsfxsize=5.5cm \epsfysize=8.0cm \epsfbox{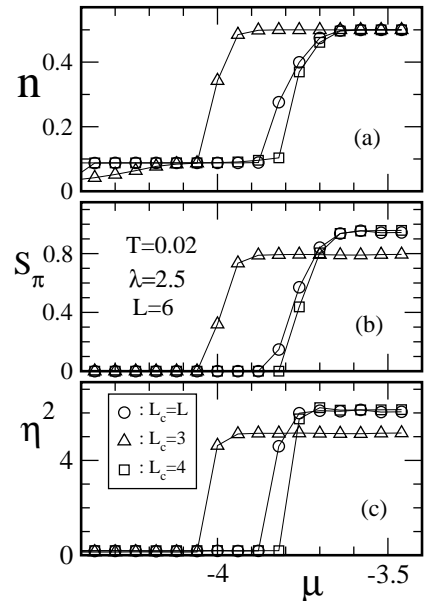}
\caption{The result of $\mu $ variation in the clean Holstein model
at coupling $\lambda=2.5$ and $T=0.02$. The exact result, with $L=6$, is
compared to results with $L_c=3$ and $L_c=4$. Panel $(a)$.~the variation
in electron density, showing the discontinuous jump from the Fermi
liquid to a charge ordered state, $(b).$~the variation in charge order
parameter with $\mu$, and $(c).$~the variance of the effective disorder
(see text)
seen by the electrons. We have not used $L_c=5$ since the results on
$L_c=4$ are already  very close to ED with $6^3$. }
\end{figure}
\end{center}

ordered
insulating (COI) phases close to $n=0.5$. The physics of
these phases has been discussed earlier 
within 
DMFT, and 
we also discuss it in detail in separate papers \cite{latt-pol1,latt-pol2}. 
Our intention here
is to estimate the effectiveness of TCA in capturing the known
features
of the Holstein model as well as compare with  exact MC calculations.

Fig.1 and Fig.2 show the variation in carrier density, $n(\mu)$, the
order parameter $S(\pi, \pi, \pi)$ for commensurate charge ordering, and
the variance of the `effective disorder', $\eta^2$ (defined further on),
seen by the electrons, with varying $\mu$ at two temperatures.
In a model which has the possibility of phase separation, and 
`disallows' a certain density range, it is imperative
to  work with constant $\mu$ to map out the phase diagram. At the 
specified $T$ and $\mu$, TCA is used to obtain a family of  equilibrium 
phonon configurations, which are then used to solve the full electron
problem. 

Fig.1 pertains to low temperature, $T=0.02$, at intermediate coupling,
where there are two phases, $(i)$~a FL at low doping, $n \lesssim 0.1$,
and $(ii)$~a commensurate COI phase for $ 0.35  \lesssim n \lesssim 0.5$,
and a regime of phase separation for $n$ between $\sim 0.1-0.35$. 
Before analysing the size dependence of the TCA results let us
define the basic indicators.
Fig.1.(a) 
shows $n(\mu)$, including the `discontinuity' due to phase
separation. Fig.1.(b) shows the COI order parameter
$S(\pi, \pi, \pi)$,
computed from the structure factor
$S({\bf q})
= (1/N^2) \sum_{ij} \langle \langle  n_i \rangle \langle n_j \rangle \rangle
e^{i {\bf q}.( {\bf R}_i - {\bf R}_j)}$,
where $\langle n_i \rangle$ is the quantum average of $n_i$ in a MC
configuration and the outer angular brackets indicate  average
over configurations.
Fig.1.(c) shows the effective disorder based on the following prescription:
for the Holstein model the electrons see a potential  
$\xi_i^{\alpha}  = \epsilon_i - \lambda x_i^{\alpha}$, where
$\epsilon_i$ is the extrinsic disorder and $x_i^{\alpha}$ is the
structural distortion in an equilibrium MC configuration ($\alpha$
is a MC configuration index).
A crude measure of the `disorder' seen by the electrons is provided by the
variance of the $\xi_i$ distribution, averaged spatially and over MC
configurations. If we denote $\eta_i = \xi_i - {\bar \xi}$, where ${\bar \xi}$
is the spatial average, then the effective disorder
$\eta^2 = \langle \eta_i^2 \rangle$. 
It  is the thermal and configuration averaged
disorder that dictates the single particle scattering, and to some extent the
transport properties. Our $\eta^2$ data in Fig.1.(c), and later figures,
quantify this disorder.

The data in Fig.1 and Fig.2 
are on a $6^3$ system, using clusters of $3^3$, $4^3$ and
$6^3$ itself to anneal the $\{ x_i \}$. Comparing the cluster size
dependence of the various physical quantities it is obvious that while
updating using the $3^3$ cluster is unable to accurately capture the effects 
in the $6^3$ system, the results based on $4^3$ are qualitatively similar
to that of $6^3$. The size difference between the cluster and the full
system is a factor of $6^3/4^3 \sim 3.4$. The ratio of the
computation time between ED-MC on $6^3$ and TCA$(6^3:4^3)$
 is $\sim 40 $, if the same extent of averaging is employed
in both calculations.

At higher temperature, $T=0.08$, Fig.2, the results based on $3^3$
continue to differ from the exact $6^3$ result
but the $4^3$ result is virtually indistinguishable from the
\begin{center}
\begin{figure}
\epsfxsize=5.5cm \epsfysize=8.0cm \epsfbox{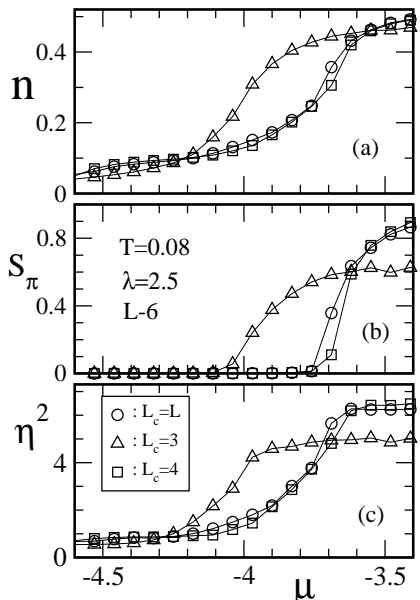}  
\caption{Same as in Fig.1, except for $T=0.08$. At this higher temperature
the coexistence jump is almost smoothed out. The agreement between
the exact result and TCA$(6^3:4^3)$ is even better here. }
\end{figure}
\end{center}
\begin{center}
\begin{figure}
\epsfxsize=5.5cm \epsfysize=8.0cm \epsfbox{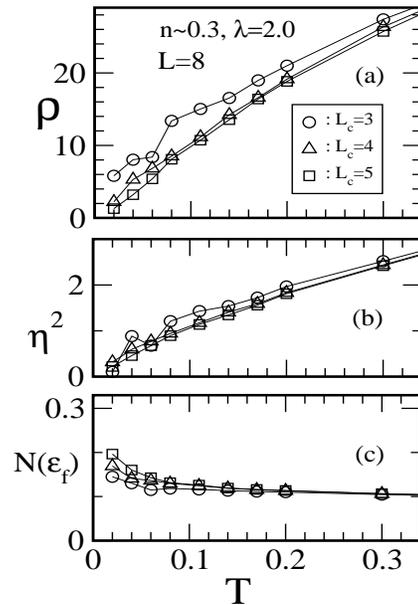}
\caption{Holstein model in the metallic regime, $\lambda=2.0$ and  $n=0.3$.
Results with TCA$(8^3:L_c^3)$, with $L_c=3-5$. Panel $(a)$: resistivity
$\rho(T)$. Except at the lowest $T$, the results with $L_c=4$ and $L_c=5$
are virtually indistinguishable, while $L_c=3$ matches the large cluster data
only at high $T$. Panel $(b)$: the effective disorder seen by the electrons, and
panel $(c)$~the density of states at the Fermi level. }
\end{figure}
\end{center}
exact answer.
The key to this lies in the large damping of the electrons arising
from thermal fluctuations, and at this temperature the effect of
the $4^3$ finite size gap is no longer relevant.

Fig.3 shows a different kind of result, where the system is studied via TCA
at
constant density, $n=0.3$, at EP coupling $\lambda=2.0$, on a ``large''
system, $L=8$, with cluster size varying from $3^3 - 5^3$. We obviously
cannot do an exact calculation on the $8^3$ system, so the 
results in Fig.3 are intended to check out 
$(i)$~the convergence of the TCA data to the $N_c=N$ 
asymptote with growing $N_c$, and $(ii)$~study the {\it temperature
dependence} of this convergence, since strong disorder, {\it i.e}, large $\eta^2$,
at high temperature could make even the $3^3$ based calculation viable.
In Fig.3 we directly compute the resistivity, using a method described
in an earlier paper \cite{sk-pm-scr}, as well as $\eta^2$ and the density
of states $N(\epsilon_F)$ at the Fermi level.

Here again, the results based on $4^3$ and $5^3$ 
clusters are virtually indistinguishable except at the lowest temperature.
The result based on $3^3$ is visibly different from that on $4^3-5^3$
at the lower temperatures, but converges to a common value for $T \sim
0.3$, by which $\eta^2$ is quite large. The results on $\eta^2$ itself
and $N(\epsilon_F)$ are quite similar for all $L_c$ at all $T$, but 
the resistivity (which is a more stringent comparison) differentiates
the changing character of the result with varying $L_c$. We would
like to draw a  general conclusion from these results, and back them
up as we discuss the disordered Holstein model in the next section.
If the single particle damping rate, $\Gamma$,
arising out of $\eta^2$ 
is comparable to the finite size gap, $12t/L_c^3$,
 in the cluster then the specific finite size features of the cluster
are smeared out and it mimics a `large' system. 
So, 
{\it annealing the variables on the large system via TCA is
effective if  $\Gamma \gg W/N_c$, where $W$ is
the bare bandwidth of the system.} For a fixed $T$ (and extrinsic 
disorder) this condition 
can be met by increasing $L_c$, while for a fixed $L_c$ the accuracy
increases as the net disorder (from thermal fluctuations and extrinsic
disorder) increases. 

To substantiate this claim, as well as check the ability of TCA to
capture the `fingerprints'
associated with a specific disorder realisation, we next consider the
disordered Holstein model.

\subsection{Disordered Holstein Model}

The key feature of the Holstein model is the possibilty of `self-trapping',
{\it i.e}, polaron formation, when the EP coupling exceeds a certain
threshold.
The critical coupling for single polaron formation \cite{pol-ref}
 in 3d is
$\lambda_c/t \sim 3.3$, which implies that the polaron `binding 
energy' $E_p^c = \lambda^2/(2K) \sim 5.44t$. This would imply that at
$\lambda=2$, where $E_p = 2t$ we should be far from any polaronic
instability. 
This is 
indeed true in the absence of disorder and Fig.3, for example,
shows that the response is metallic with $d\rho/dT >0$. However,
even weak disorder, $\Delta=0.6$, has dramatic effect in the FL
phase, see Fig.4. This figure presents results on 
 the disordered Holstein
model studied 
directly via ED-MC on $6^3$, as well as by TCA using
$3^3$ and $4^3$ clusters on the $6^3$ system.

\begin{center}
\begin{figure}
\epsfxsize=6.0cm \epsfysize=8.0cm \epsfbox{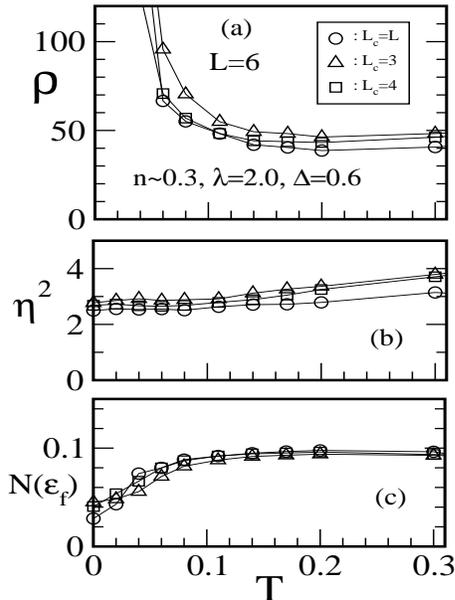}
\caption{Disordered Holstein: $\lambda=2.0$, $\Delta=0.6$,  and  $n=0.3$,
exact results using $6^3$ and TCA.
TCA is based on  $L_c=3$ and $L_c=4$. Panel $(a).$~resistivity
$\rho(T)$,  panel $(b).$~the effective disorder $\eta^2$, 
panel $(c).$~the density of states at the Fermi level. }
\end{figure}
\end{center}

\begin{center}
\begin{figure}
\epsfxsize=6.5cm \epsfysize=6.5cm \epsfbox{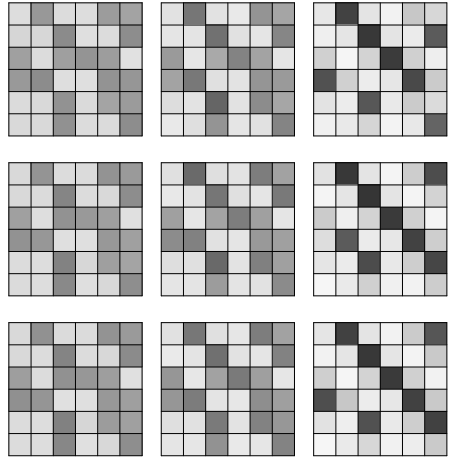}
\caption{Thermally averaged density profile, top surface of a 
$6^3$ system: 
comparing exact results on $6^3$ with TCA.
The top row is for $L_c=3$, next for $L_c=4$ and lowest for $L_c=L=6$.
Temperature, left to right, are $0.3,~0.11,~0.01$. 
The disorder realisation $\{\epsilon_i\}$ is the same in all cases.
 }
\end{figure}
\end{center}

In terms of the physical effect of disorder, the ED-MC indicates that
the interplay of disorder and EP coupling can turn the system into
a very bad metal (or even  insulator)
with a large resistivity at $T=0$,
and $d\rho/dT <0$ for $T \rightarrow 0$, Fig.4.(a).
The 
`effective disorder' seen by the electrons is large
 down to $T=0$,  Fig.4.(b), and there is a pseudogap
in the DOS, as evident from $N(\epsilon_F)$, Fig.4.(c).

The conversion of a FL (at $\lambda=2.0, \Delta=0$) into a
`polaronic' phase by weak disorder happens because the
density inhomogeneity created by weak disorder is 
dramatically amplified by strong EP coupling \cite{latt-pol2,emin-bussac}
leading to strong localisation. However, all electronic states are
not strongly localised, as the spatial pattern, Fig.5, and $N(\epsilon_F)$,
Fig.4.(c), indicate.
Fig.5, discussed further on,
shows the thermally averaged density $n_{\bf r}$.

In contrast to our results on the clean Holstein model, Fig.1-3, notice 
that {\it all} 
the sizes, $3^3$, $4^3$ and $6^3$,
 yield similar results on all the indicators, Fig.4.(a)-(c). While
the $4^3$ based results almost coincide with the exact $6^3$ answer
even the $3^3$ based results capture all the qualitative features and
even the numerical values reasonably accurately. The large effective 
disorder in the problem, arising from the strong lattice distortions,
$x_i$, makes even the $3^3$ calculation acceptable. The $\eta^2$ in this
model is $\sim 3.0$ over the whole $T$ range, comparable to the {\it maximum}
$\eta^2$ in the clean problem, Fig.3.(b).

While the transport and spectral properties seem to be adequately captured 
by TCA, does the method succeed in capturing the specific `fingerprint' of
a disorder realisation, $\{ \epsilon_i \}$? Fig.5 shows the thermally averaged
density pattern, $n_{\bf r}$, at three different temperatures (along the row)
computed via TCA using $L_c=3$ (first row), $L_c=4$ (second row) and the 
exact, $L_c=6$ case (bottom row). The TCA was run with the {\it same 
realisation} of disorder in all three cases. Remember that while the cluster
based update is used for the phonon degrees of freedom, the final density field
is calculated by diagonalising the {\it full Hamiltonian} in the background
of the quenched disorder and the phonon configurations obtained via TCA. The
cluster diagonalisation by itself cannot yield the density field.

At first glance, the results of all three runs, compared along a column, match
 well. At intermediate and high  temperature there is weak but 
still visible
density contrast and the results of all three schemes match very well. At the
lowest temperature, the contrast is strongest and, although the correspondence
along the third column  is
quite striking, there are some minor variations between the panels.
This is partly because the MC based annealing is less effective at
low temperature due to the small acceptance rate of moves. 
Apart from this generic difficulty with MC calculations, we think
the 
overall
ability of TCA to capture the specific features of a disorder realisation
(and not just system averaged properties) is quite impressive.

Now consider using TCA on large sizes, $L=8$, as in Fig.3, 
for the disordered problem. Fig.6 shows TCA based results using $L_c=3, 4, 5$
to solve the $L=8$ problem, and Fig.7 shows the associated density profile.
As stated before it is impossible to  do  ED-MC on $8^3$, so the data here
is  meant to indicate the convergence of the  TCA based results to the 
$N_c \rightarrow N$
asymptote, although the exact result at $N_c=N$ is not available.
The effective disorder, Fig.6.(b), and $N(\epsilon_F)$, Fig.6.(c), are
very similar for all $L_c$, and even the resistivity, Fig.6.(a), 
matches quite well after disorder average over $\sim 4-5$ copies.

\begin{center}
\begin{figure}
\epsfxsize=6.0cm \epsfysize=7.5cm \epsfbox{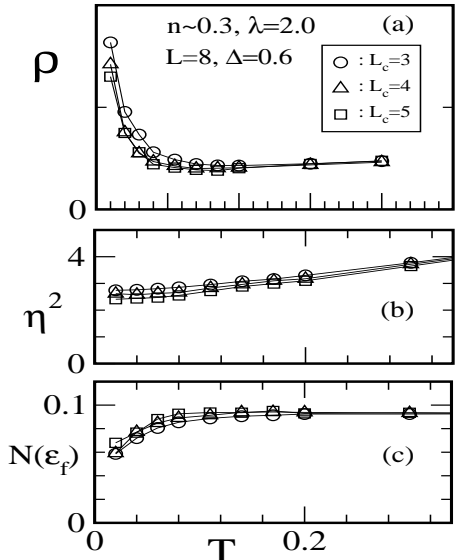}
\caption{Disordered  Holstein model, $\lambda=2.0$ and  $n=0.3$,
$\Delta=0.6$.
Results with TCA$(8^3:L_c^3)$, with $L_c=3-5$. Panel $(a)$: resistivity
$\rho(T)$, 
$(b)$: the effective disorder seen by the electrons, and
$(c)$~the density of states at the Fermi level. }
\end{figure}
\end{center}

\begin{center}
\begin{figure}
\epsfxsize=7.0cm \epsfysize=7.0cm \epsfbox{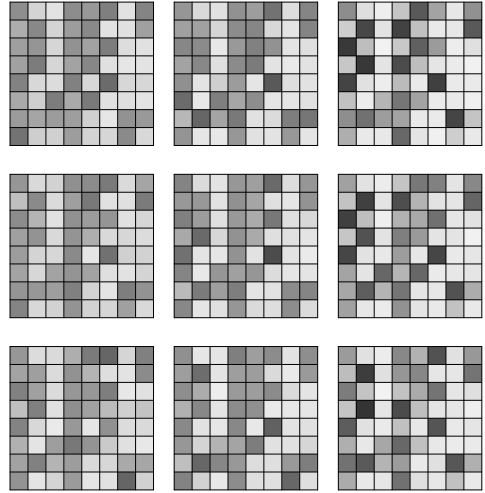}
\caption{Thermally averaged density profile: 
comparing TCA based results on a $8^3$ system using $L_c =3,4,5$.
The top row is for $L_c=3$, next for $L_c=4$ and lowest for $L_c=L=5$.
Temperature, left to right, are $0.4,~0.14,~0.04$. 
The disorder realisation $\{\epsilon_i\}$ is the same in all cases.}
\end{figure}
\end{center}

We would argue that for
the conditions studied, TCA with $L_c=3-5$ is quite adequate in annealing the
phonon variables on $L=8$. {\it In fact when $L_c$ is large enough, 
so that $\Gamma \gtrsim 12t/L_c^3$, the outer
limit, i.e, system size $L$, is actually irrelevant}. 
Updates based on small 
clusters can successfully generate the appropriate configurations on large
lattices. The only reason large $L$ is needed at all, for the system as
a whole, is to compute transport
properties, or check for long range order.

The thermally averaged 
density profile corresponding to the MC in Fig.6 are shown in Fig.7.
As in Fig.5, the patterns along each row correspond to decreasing $T$,
reducing from $T=0.4$ to $0.14$ to $0.04$.
The first row corresponds to TCA with $L_c=3$, the second with $L_c=4$ and
the third to $L_c=5$. All three systems have the same realisation of
quenched disorder $\{\epsilon_i\}$. Again, as in Fig.5, the different TCA
results are in excellent agreement at higher temperature (first two columns)
while there are some differences at the lowest temperature due to the 
difficulty in annealing.  The actual percent difference in the density field,
between the three systems, averaged over the whole $8^3$ system is $\sim 10-15\%$.

\subsection{Double Exchange Model}

The clean double exchange model has been widely studied, using a 
variety of analytical approximations and numerical techniques. We 
do not enter into a detailed recapitulation of these results since
we have already reviewed them in detail \cite{sk-pm-scr} in an
earlier paper. Here  we focus primarily on MC based results, since these
are unbiased, although necessarily finite size.
The ground state of the clean DE model is a saturated ferromagnet
and in terms of transport both the ferromagnetic and paramagnetic 
phase are metallic.

In the limit $J_H/t \rightarrow \infty$ the DE model 
maps on to 
a spinless fermion problem, with `hopping disorder' arising from
the background spin configuration. Since the strong local coupling $J_H$
couples the core spin orientation to fermion spin projection, only the
`locally aligned' fermion state is viable at each site (the other is
at an energy $J_H$ above). The electron {\it hopping} between two sites 
depends on the electronic eigenfunctions at each site, and so on the
spin orientation.  
The `projected' Hamiltonian \cite{sk-pm-scr} 
turns out to be:
\begin{eqnarray}
H & =&
-t\sum_{\langle ij \rangle}
(~g_{ij}  \gamma^{\dagger}_i  \gamma_j +
h.c~) 
-\mu \sum_i  n_i \cr
&=& -t\sum_{\langle ij \rangle} f_{ij}
(~e^{i \Phi_{ij}}  \gamma^{\dagger}_i  \gamma_j +
h.c~) 
-\mu \sum_i  n_i
\end{eqnarray}
The $\gamma$'s are spinless fermion operators.
The hopping amplitude $g_{ij} = f_{ij} e^{i\Phi_{ij}}$
between locally aligned states,
can be written in terms of the polar angle $(\theta_i)$ and
azimuthal angle $(\phi_i)$ of the spin ${\bf S}_i$
as,
$  cos{\theta_i \over 2} cos{\theta_j \over 2}$
$+
sin{\theta_i \over 2} sin{\theta_j \over 2}
e^{-i~(\phi_i - \phi_j)}$.
It is easily checked that
the `magnitude'  of the overlap,
$f_{ij} = \sqrt{( 1 + {\bf S}_i.{\bf S}_j)/2 }$,
while the phase is specified by
$tan{\Phi_{ij}} = Im(g_{ij})/Re(g_{ij})$.

Fig.8 shows MC results on the DE model, using $L_c=4$ and 
varying the system size from $L=4$ to $L=10$. This is unlike
our earlier results on the Holstein model where we kept the
system size fixed and varied $L_c$, looking for convergence.
For the DE model, there is already  MC data available \cite{furu-clean}
on large sizes.  Fig.8 shows the evolution of the 
magnetisation profile $m(T)$ as $L$ increases. 

\begin{center}
\vspace{.6cm}
\begin{figure}
\epsfxsize=6.0cm \epsfysize=5.5cm \epsfbox{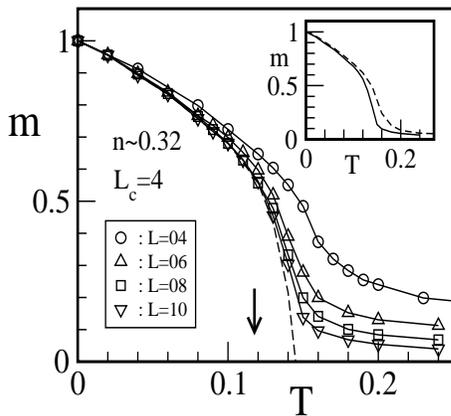}
\caption{Magnetisation in the clean DE model with $L_c =4$, and $L=4-10$.
The electron density is $n=0.32$.  The magnetisation profile
obtained by extrapolation to $L\rightarrow \infty$ is shown as a
dotted line. 
The arrow indicates $T_c$ using the 
the moment expansion based MC and finite size scaling by 
Furukawa {\it et al.}. The inset compares the TCA result to
an earlier `effective Hamiltonian' approximation made
by us. }
\end{figure}
\end{center}

\begin{center}
\begin{figure}
\epsfxsize=5.2cm \epsfysize=8.0cm \epsfbox{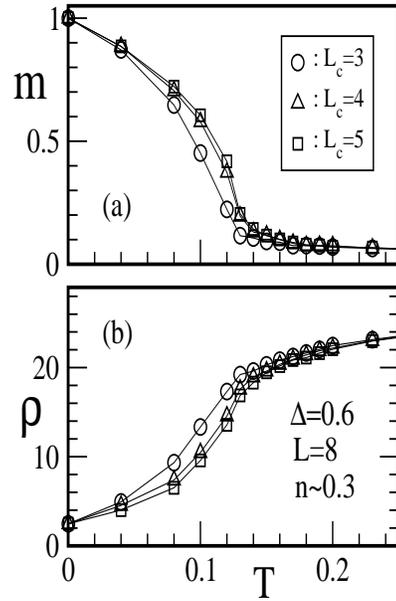}
\caption{Disordered  DE model: magnetisation and resistivity.
with $L_c =3,4,5$, and $L=8$.
The electron density is $n=0.30$.}
\end{figure}
\end{center}

The large $L$
extrapolation of this trend, at $L_c=4$, suggests $T_c \approx 0.14$.
The result of moment expansion based MC and finite size scaling \cite{furu-clean}
indicate $T_c \approx 0.12$ at this density. Although there is a 
difference between these results, the ability of TCA to approach the
exact answer, with substantially less computational effort, is
evident. The inset in Fig.8 compares an earlier approximation
used by us \cite{sk-pm-scr}
 (using a classical effective Hamiltonian for the spins)
with the TCA using $L=10, L_c=4$. The improvement in TCA, particularly
in capturing the $T_c$ scale is obvious.

Fig.9 shows results on the DE model in the presence of weak disorder in
the electron system. This model has an additional term $\sum_i \epsilon_i n_i$
where the $\epsilon_i$ is binary disorder with strength $\pm \Delta$ as in
the Holstein problem. The data on magnetisation and resistivity in
Fig.9 is shown for $N=8^3$, with $L_c=3, 4, 5$. As in the Holstein
problem, the profiles with $L_c=4$ and $L_c=5$ are barely 
distinguishable.

\subsection{Computational indicators}

\subsubsection{Equilibriation}

Since the TCA based MC does not compute the total energy of the system 
in the process of updating, unlike ED-MC, the process of equilibriation
and stability of the energy is not obvious. To explicitly confirm the 
nature of equilibrium fluctuations (and the absence of drift in the mean
value) as well as visualise the process of equilibriation in response to
a temperature step, we ran TCA with a simultaneous calculation of the 
total energy at the end of each MC sweep. The results, for the Holstein 
model, 

\begin{center}
\vspace{.5cm}
\begin{figure}
\epsfxsize=7.0cm \epsfysize=5.0cm \epsfbox{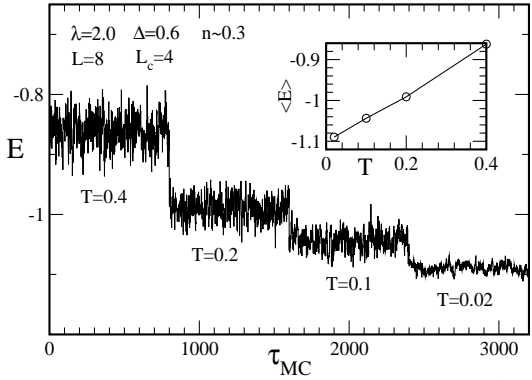}
\caption{Total energy variations during the TCA-MC on the 
Holstein model. The electronic parameters and temperatures are
 marked on the
Figure.   This trace is quite similar to that obtained in a typical
ED-MC run, and the energy fluctuations reduce as expected with
decreasing temperature. The inset shows the thermally averaged 
energy at each $T$ as a function of temperature.}
\end{figure}
\end{center}

are shown in Fig.10. Notice that this requires diagonalisation
of the full Hamiltonian matrix $N_{MC}$ times, rather than $N_{MC}N$
times as required by ED-MC. Apart from assurance about proper
equilibriation
this diagnostic allows us to track potential hysteresis effects,
{\it i.e}, difference between heating and cooling, if a first order
transition is involved.

\subsubsection{Comparing $\Delta {\cal E}$ }

It may be useful to check out the claim that
{\it microscopically} the TCA based calculation  of $\Delta {\cal E}$
quickly converges with increasing $N_c$ (even if it is still $\ll N$)
and with increasing extrinsic disorder.
To that effect Fig.11 analyses the results on the Holstein model.
We update the system using TCA with $L_c=3-5$ on a $L=8$ system.
For each $L_c$, and specified $T$, we choose a reference site
randomly, and whenever a phonon update is attempted at that site in
the course of the MC sweep, we not only compute the cluster based
energy difference $\Delta {\cal E}_c$ but also the {\it exact}
energy cost of such a move $\Delta {\cal E}_{ED}$ based on the
full system. 

The system evolves accoring to TCA, but we keep track
of these energy differences (at that site) and construct the 
following error measure:
\begin{equation}
\delta(L_c, L) = {1 \over N_{MC}} \vert
{ {\Delta {\cal E}_c - \Delta {\cal E}_{ED} } \over \Delta {\cal E}_{ED}}
\vert
\end{equation}
where the averaging is over MC steps, and the `error' $\delta$ implicitly
depends on temperature,  as well as all other electronic parameters (in
particular, disorder).

Fig.11 shows results, again on the Holstein model, with $L=8$ and
$L_c=4$, panel $(a)$. The state at $T=0$ is a {\it clean Fermi liquid}
for $T=0$, the electronic states are simple tight-binding states and,
as expected, $\Delta {\cal E}$ computed on $8^3$ and $4^3$ have a fair
difference, 
about $25 \%$. This ``error'' 
% ---------------------------------------------
\begin{center}
\vspace{.5cm}
\begin{figure}
\epsfxsize=7.0cm \epsfysize=7.0cm \epsfbox{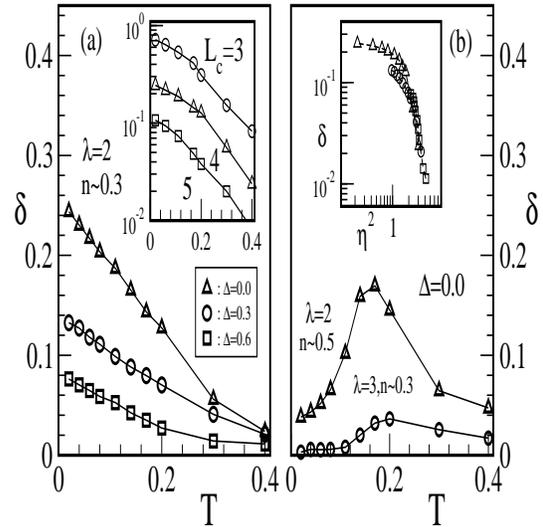}
\caption{Error in energy estimate of MC moves. The indicator $\delta$
defined in the text, compares energy costs on the full $L^3$ system
with that based on $L_c^3$ clusters. Panel $(a)$~shows the $T$ 
dependence of $\delta$ with increasing disorder, and the inset
shows the $L_c$ dependence at $\Delta=0$. Panel $(b)$~same as
in panel $(a)$ with different parameters, marked on the Figure.
Inset to panel $(b)$~variation of $\delta$ with the effective disorder,
$\eta^2$, combining the results in main panel $(a)$.
}
\end{figure}
\end{center}
% ---------------------------------------------
falls quickly 
with increasing temperature, as thermal fluctuations in the $x_i$ 
increase (staying at $\Delta=0$), 
and also reduces systematically on introduction
of even weak disorder $\Delta =0.3-0.6$.
The inset to Fig.11.$(a)$ shows temperature  dependence of the error
with varying $L_c$ at $\Delta=0$. 
To construct an approximate measure, the typical error, averaged over the
temperature window is $\sim 0.5$ at $L_c=3$, $\sim 0.1$ at $L_c=4$,
and $\sim 0.05$ at $L_c=5$. 
Fig.11.$(b)$ shows $\delta$ at $n=0.5, \lambda=2.0$ (which is a charge
ordered state at low temperature) and $n=0.3, \lambda=3.0$ (which is
a polaronic insulator) again at $L=8$ with $L_c=4$. The error in the
CO problem is non monotonic because, unlike the Fermi liquid, the
system goes into a ordered chessboard phase at $T=0$ and this 
localisation reduces the error. In the polaronic insulator
phase the error is below the $5\%$ threshold at all temperatures
due to the strongly localised nature of electronic wavefunctions.
The inset to Fig.11.$(b)$ puts together the error variation with
respect to the effective disorder, $\eta^2$, for the data in
Fig.11.$(a)$ for $\Delta=0-0.6$ with
varying temperature. As we have argued earlier there is a roughly
`universal' behaviour of the error in terms of the effective disorder,
{\it irrespective of its origin},`
and at $L_c=4$ 
an error $\lesssim 0.1$ is obtained whenever $\eta^2 \gtrsim 1.0$

Note that $\delta$ above is a measure of error in the energy estimate,
we have checked that the error in $e^{-\beta \Delta {\cal E}}$ 
itself, which controls the MC moves,  between
the $4^3$ and $8^3$ cluster are much 
smaller than the error in $\Delta {\cal E}$.
This 
explains why the thermodynamic and transport results are
accurate even with $L_c=4$ in the clean system.

\section{Conclusion }

Since we have put forward the TCA based approach to adiabatic problems
as a `general' many body technique, not restricted to the specific 
examples studied here, let us compare it with DMFT \cite{dmft-ref}
 which finds wide
use as a method for handling correlated electrons.
$(i)$~DMFT maps on a correlated  lattice problem to an impurity model, 
reducing it to an effective single site problem and making it
more tractable. 
The adiabatic approach, and the TCA approximation,
ignores the quantum 
dynamics of the background fields, leading to a 
static but `annealed' disorder
problem. 
$(ii)$~The two methods have complementary reach. DMFT is good at handling
local quantum correlations but misses out on spatial fluctuations and
disorder physics. The real space adiabatic approach, supplemented by
TCA like approximations, handles disorder and spatial correlations
accurately but cannot  handle
interacting quantum modes.
$(iii)$~Both have had significant success in materials physics. DMFT
extended to include non local correlations 
will be truly global method, the 
adiabatic approach extended to include quantum dynamics,
and implemented via TCA like approximations, or refinements,
would `reach' real life problems from another direction.

In conclusion, we have put forward and benchmarked a new Monte
Carlo technique for handling fermions strongly coupled to
classical degrees of freedom. The method relies on the approximate 
`locality'
of the energy cost of a MC move in a system with moderate disorder,
allowing accurate estimate of energy differences to be made 
using on a cluster Hamiltonian instead of the full system. 
This allows annealing of
the classical variables on large lattices, breaking the $N^4$ 
barrier that plagues exact diagonalisation based Monte Carlo.
The approach makes no assumptions regarding the  
the starting Hamiltonian, except the quadratic nature of
the quantum degrees of freedom. We have provided 
basic benchmarks on
the Holstein and Double Exchange model, and will provide 
several new results in forthcoming papers.

\vspace{.2cm}

We acknowledge use of the Beowulf cluster at HRI.


\begin{thebibliography}{99}
\bibitem{dmrg-ref}
S. R. White, Phys. Rev. Lett. {\bf 69}, 2863 (1992), 
see Karen Hallberg, cond-mat 0303557, for a recent review. 
\bibitem{dmft-ref} 
Antoine Georges, Gabriel Kotliar, Werner Krauth and Marcelo J. Rozenberg,
Rev. Mod. Phys. {\bf 68}, 13 (1996).  For recent reviews see,
Tudor Stanescu and  Gabriel Kotliar, cond-mat 0404722,
Antoine Georges, cond-mat 0403123.
\bibitem{adiab-elphon} 
A.B. Migdal, Sov. Phys. JETP {\bf 7}, 996 (1958),
V.V. Kabanov and O.Y. Mashtakov, Phys. Rev. {\bf B 47}, 6060 (1993).
\bibitem{car-parin}
R. Car and M. Parrinello, Phys. Rev. Lett. {\bf 55}, 2471 (1985).
\bibitem{mill-mull}
A. J. Millis, B. I. Shraiman, and R. Mueller, Phys. Rev. Lett. {\bf 77}, 
175 (1996). 
\bibitem{dag-ref} See {\it e.g},
S. Yunoki, J. Hu, A. L. Malvezzi, A. Moreo, N. Furukawa, and E. Dagotto,
Phys. Rev. Lett. {\bf 80}, 845 (1998), S. Yunoki, T. Hotta, and E. Dagotto,
Phys. Rev. Lett. {\bf 84}, 3714 (2000). 
\bibitem{de-mc}
M. J. Calderon and L. Brey, Phys. Rev. {\bf B 58},
3286 (1998), Y. Motome and N. Furukawa, J. Phys. Soc. Jpn.
{\bf 68}, 3853 (1999).
\bibitem{dms-refs} 
G. Alvarez, M. Mayr, and E. Dagotto, Phys. Rev. Lett. 
{\bf 89}, 277202 (2002). 
\bibitem{sk-pm-dde} Sanjeev Kumar and Pinaki Majumdar,
Phys. Rev. Lett. {\bf 91}, 246602-1 (2003).
\bibitem{sk-pm-nano} Sanjeev Kumar and Pinaki Majumdar,
Phys. Rev. Lett. {\bf 92}, 126602 (2004).
\bibitem{ed-mc-ref} 
S. Yunoki, {\it et al.}, Phys. Rev. Lett. {\bf 80},
845 (1998), E. Dagotto, {\it et al.}, Phys. Rev. {\bf B 58}, 6414 (1998).
\bibitem{furu-clean} Yukitoshi Motome and  Nobuo Furukawa, 
J. Phys. Soc. Jpn. {\bf 72}, 2126 (2003).
\bibitem{furu-dis}
Y. Motome and N. Furukawa, J. Phys. Chem. Solids,
{\bf 63}, 1357 (2002). 
\bibitem{hyb-mc} 
J. L. Alonso, L. A. Fernandez, F. Guinea, V. Laliena, V. Martin-Mayor,
Nucl.Phys. {\bf B 596}, (2001) 587-610
\bibitem{sk-pm-scr} Sanjeev Kumar and Pinaki Majumdar, cond-mat 0305345.
\bibitem{dmft-holst1} A. J. Millis, R. Mueller, and B. I. Shraiman,
Phys. Rev. {\bf B 54}, 5389 (1996).
\bibitem{dmft-holst2} S.  Ciuchi and F. de Pasquale,
Phys. Rev. {\bf  B 59}, 5431 (1999).
\bibitem{dmft-holst3} S. Blawid and A. J. Millis, Phys. Rev. {\bf B 62},
2424 (2000).
\bibitem{pol-ref} See, {\it e.g}, A. H. Romero, D. W. Brown and K. Lindenberg,
Phys. Rev. {\bf B 60}, 14080 (1999).
\bibitem{latt-pol1} Sanjeev Kumar and Pinaki Majumdar, cond-mat 0406083. 
\bibitem{latt-pol2} Sanjeev Kumar and Pinaki Majumdar, cond-mat 0406084. 
\bibitem{emin-bussac} See, {\it e.g}, D. Emin and M.-N. 
Bussac, Phys. Rev. {\bf B 49},
14290 (1994) for the interplay of {\it extrinsic} disorder and EP coupling.
\end{thebibliography}
\end{document}